\documentclass{elsart}
\usepackage{graphicx,amssymb,amsmath}

\font\FermiPPTfont=cmssbx10 scaled 1440
\font\FermiSmallfont=cmssq8 scaled 1200

\journal{Astroparticle Physics}

\def\ob{\Omega_b h^2}

\def\FNALppthead#1#2#3{
\null 
\begin{center}
\vskip -1.0truein{\hspace*{-2truecm}\hbox to 18.5truecm {
\vbox to 1in{\vfill 
             \hbox{\includegraphics[height=1.5cm]{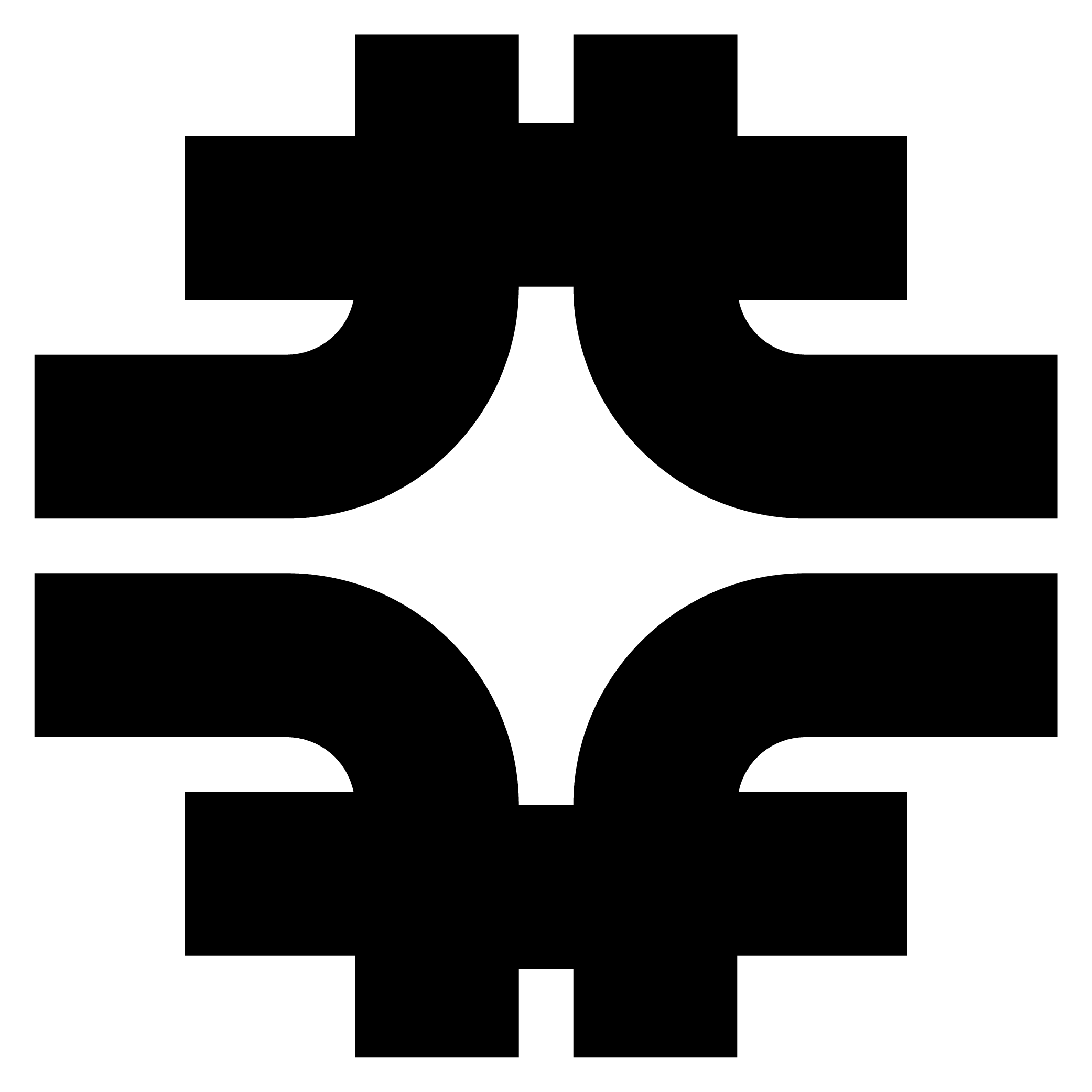}}
             \vfill }
\hskip 1em
\vbox to 1in{\vfill
             \hbox{{\FermiPPTfont Fermi National }}
             \hbox{{\FermiPPTfont Accelerator Laboratory}}
             \vfill}
\hfill
\vbox to 1in {\vfill \FermiSmallfont
              \hbox{#1}
              \hbox{#2}
              \hbox{#3}
              \vfill}
}}\vskip-0.0truein\end{center}}

\begin{document}
\FNALppthead{NASA/Fermilab Astrophysics
  Center}{astro-ph/0205238}{FERMILAB-Pub-02/083-A} 

\begin{frontmatter}
\title{Telling Three from Four Neutrinos\\ with Cosmology}
\author{Kevork N. Abazajian}
\ead{aba@fnal.gov}
\address{NASA/Fermilab Astrophysics Center, 
Fermi National Accelerator Laboratory, Batavia, Illinois 60510-0500,
USA}
\date{7 August 2002}

\begin{abstract}
New results, namely the independent determination of the deuterium
abundance in several quasar absorption systems, and the complementary
determination of the cosmological baryon density by observations of
anisotropies in the cosmic microwave background (CMB), allow for a
reevaluation of the constraints on the relativistic particle content
of the universe at primordial nucleosynthesis.  Expressed in terms of
the neutrino energy density, we find $1.7 < N_\nu < 3.5\
(95\%\rm~CL)$.  In particular, we show that phenomenological four
neutrino models including a sterile state (not participating in
$SU(2)_L\times U(1)_Y$ interactions) unavoidably thermalize a fourth
neutrino, and are highly disfavored in the standard minimal model of
primordial nucleosynthesis, if the systematic uncertainty in the
primordial helium abundance is small.  We describe plausible
extensions of the minimal model which evade this constraint.
\end{abstract}



\end{frontmatter}

\section{Introduction}

With the simplifying conditions of isotropy, homogeneity, thermal
equilibrium, and the particle content of the standard model of
particle physics, big bang nucleosynthesis (BBN) is a one parameter
model dependent only on the baryon-to-photon ratio, $\eta$, at that
epoch.  The appeal of this simple yet successful standard
model~\cite{bbnreview1} has motivated a predictive ability, for
example, in the number of leptons in the standard model of particle
physics~\cite{Steigman:kc}.

The light nuclides D, $^3$He, $^4$He and $^7$Li are produced in
measurable quantities in the first three minutes of the standard
cosmology.  Considerable attention has been devoted to the analysis of
uncertainties in the predicted abundances of the light elements and
their consistency with the observed light element
abundances~\cite{bbnconsist,Burles:2000zk,Cyburt:2001pq,rpp}.  Though
systematic uncertainties likely dominate the helium abundance
measurements~\cite{rpp,sauerjedamzik,os} and may be present in
observations of the deuterium-to-hydrogen ratio D/H in high-redshift
quasar systems~\cite{omeara}, and $^7$Li may be partially depleted by
stellar processes~\cite{lideplete}, standard BBN is remarkably
successful predictor for the abundances of these light elements with
abundances that differ by nine orders of magnitude.

There exist a variety of ways of modifying the standard BBN paradigm,
including altering the spatial distribution of baryon number,
out-of-equilibrium decays of massive particles, or new neutrino
physics (for a summary, see Ref.~\cite{karsten}).

We focus our attention on a minimal extension to BBN by a modification
of the neutrino sector, and specifically to models which attempt to
simultaneously account for the indications of neutrino mixing and
masses from the atmospheric neutrino results of
Super-Kamiokande~\cite{superk}, the observations of the transformed
solar neutrino flux~\cite{solar,sno}, and the Liquid Scintillator
Neutrino Detector (LSND) signal~\cite{lsnd}.  A class of models that
can accommodate all of these results introduce a fourth mass
eigenstate~\cite{maltoni}. As is well known, the fourth flavor state
must be sterile, {\it i.e.}, not participating in $SU(2)_L\times
U(1)_Y$ interactions, due to its being both light ($m \ll 1\rm\ GeV$)
and not observed in the invisible width of the $Z^0$ boson~\cite{rpp}.
In current manifestations of four neutrino mixing models, the sterile
neutrino is not necessarily closely associated with a single mass
eigenstate, since the atmospheric and solar observations each disfavor
large sterile components.  However, recent global
analyses~\cite{maltoni} of the available neutrino oscillation data,
including short baseline limits~\cite{choozpaloverde}, leave
four-neutrino models viable.  There are several ways that light
sterile neutrinos can be accommodated in neutrino mass models.  For a
review see, {\it e.g.}, Ref.~\cite{Volkas:2001zb}.

The standard contribution to the energy density by the three active
neutrinos can be augmented by the complete or partial equilibration of
the sterile mode.  The increased energy density in units of the energy
density in one neutrino and its antiparticle ($\rho_{\bar\nu\nu} =
7\pi^2 T^4/120$) is then $N_\nu = 3 + \Delta N_\nu$, where $\Delta
N_\nu = \rho_s/\rho_{\bar\nu\nu}$ is the relative contribution of the
sterile state.  Here, we use $\Delta N_\nu$ strictly as a
parameterization of extra (or missing) relativistic energy density.
Sterile neutrinos in the early universe can also give rise to lepton
asymmetry generation~\cite{leptasym}, which can alter or strongly
suppress sterile neutrino thermalization, or, if the asymmetry is
generated in the $\nu_e/\bar\nu_e$ sector, alter beta-equilibrium and
thus light element abundance production, primarily in the production
of $^4$He.  As an alternate measure, the baryon-to-photon ratio $\eta$
is related to the cosmological baryon density $\Omega_b$ (as a
fraction of the cosmological critical density) as $\eta \simeq
2.74\times 10^{-8}\, \ob$, where $h$ is the present Hubble parameter
in units of $100\rm\ km\ s^{-1}\ Mpc^{-1}$.

Letting $N_\nu$ be a free parameter of BBN, its primary effect is
altering the predicted $^4$He abundance $Y_p$.  The remaining
parameter, the baryon content $\eta$, is (over) constrained by D/H,
$^3$He and $^7$Li.  In one analysis, Lisi, Sarkar and
Villante~\cite{lisi} used four permutations of primordial light
element abundance determinations to derive limits roughly in the range
$2<N_\nu < 4$.  In one combination of light element abundance
determinations (their data set A), the 99.7\% CL region allowed $N_\nu
\sim 4.5$.  This value is widely cited as allowing for an additional
neutrino (or relativistic degree of freedom) at BBN.  Though this
limit was correct, it relied on the possibility of a ``high''
primordial deuterium abundance.  Since that work, deuterium has been
observed or bounded to have a ``low'' value in six high-redshift
quasar absorption systems (QAS) by three groups~\cite{omeara,sixdh},
and the ``high'' deuterium QAS observation~\cite{webb} has not been
verified in other systems and is disputed~\cite{highbye}.  Using the
low deuterium abundance and a small uncertainty in $Y_p$, Burles {\it
et al.}~\cite{Burles:1999zt} found the limit $N_\nu < 3.20$ with the
prior $N_\nu > 3$. The work by Cyburt, Fields \&
Olive~\cite{Cyburt:2001pq} took the possibility of two values of the
baryon density, that given by D/H+BBN and CMB, $\Omega_b h^2 \simeq
0.02$, and a value of $\ob \simeq 0.01$ preferred by one inferred
primordial value of $Y_p$~\cite{oss} and the $^7$Li
abundance~\cite{ryan} (if undepleted).  Ref.~\cite{Cyburt:2001pq}
finds that these densities give 95\% CL upper bounds of $N_\nu < 3.6$
and $N_\nu < 3.9$, respectively.

Motivated by the six quasar absorption system measurements of a
``low'' D/H and the analysis of the observations of the CMB anisotropy
experiments DASI, BOOMERanG, and MAXIMA, we adopt that the inferred
value of the cosmic baryon density from D/H plus standard BBN $\ob
(\rm D/H)$ and the shape of the acoustic peaks in the CMB angular
power spectrum $\ob ({\rm CMB})$, are approaching the actual value of
$\ob$, within statistical and systematic uncertainties.  The
forthcoming analysis of the Microwave Anisotropy Probe (MAP)
satellite's observations will potentially reduce the uncertainty in
$\ob$ to approximately 10\%~\cite{map}.  In Section~\ref{bbn}, we
analyze in detail the constraints arising from accurate calculations
of the primordial helium abundance, the inferred primordial helium
abundance, using either $\ob (\rm D/H)$ or $\ob ({\rm CMB})$,
and show that a thermalized fourth neutrino is highly disfavored by
standard BBN.  In Section~\ref{neutrino}, we analyze current four
neutrino mixing schemes and their behavior in the early universe and
show that thermalization of a fourth neutrino state is unavoidable.
In Section~\ref{caveat}, we describe various means and methods of
extending the standard model to evade this constraint.

\section{The BBN Prediction of the Number of Neutrinos}
\label{bbn}
The consistency of BBN as a predictor of the light element abundances
already been explored in some
detail~\cite{bbnconsist,Burles:2000zk,Cyburt:2001pq,rpp}.  We instead
focus on the current uncertainties in the cosmic baryon density $\ob$
and the observed primordial $^4$He abundance $Y_p$, the light nuclide
whose abundance is most sensitive to the energy content of the
universe at the BBN epoch.

The baryon content of the universe can be estimated in a variety of
ways, of which the most precise measures currently are the deuterium
abundance at high redshift and the shape of the acoustic peaks in the
CMB~\cite{newsarkar}.  Deuterium has been observed in high-redshift
metal-poor neutral hydrogen systems which are seen as absorbers in the
spectrum of back-lighting quasars.  The deuterium in these extremely
metal-poor systems is inferred to be close to the primordial value due
to minimal stellar processing which produces metals and only destroys
deuterium.  Due to the extreme sensitivity of D/H to the baryon
content at BBN, the baryon density required to produce the observed
deuterium abundance is rather precisely
determined~\cite{Burles:2000zk,omeara}:
\begin{equation}
\label{omegabdh}
\ob({\rm D/H}) = 0.020\pm 0.002\ ({\rm 95\% CL}),
\end{equation}
for the standard energy density content $N_\nu=3$.  The baryon density
inferred from D/H has a small dependence on the relativistic energy
density of the plasma, which is~\cite{Cardall:1996ec}
\begin{equation}
\label{omegabdhnnu}
\ob({\rm D/H}, N_\nu) = \ob({\rm D/H}, N_\nu=3)(1+0.125\,\Delta N_\nu).
\end{equation}

The baryon content also alters the amplitude of acoustic oscillations
in the primordial plasma at CMB decoupling and the relative height of
the first three acoustic peaks (for a review of the physics of the
CMB, see Ref.~\cite{hudodelson}).  The first three acoustic peaks in
the angular power spectrum of the CMB have been detected in the
analysis of CMB anisotropy measurements by the DASI~\cite{dasi},
BOOMERanG~\cite{boom} and MAXIMA~\cite{maxima} experiments.  The
results of these experiments' analyses find
\begin{align}
\ob\,({\rm D}) &= 0.022^{+0.004}_{-0.003}\ &({\rm 95\% CL})
\label{dasiomegab} \cr
\ob\,({\rm B}) &= 0.022^{+0.004}_{-0.003}\ &({\rm 95\% CL})\cr
\ob({\rm M}) &= 0.033\pm 0.013\ &({\rm 68\% CL})
\end{align}
respectively.  The BOOMERanG value above is that given by their
Bayesian approach.  For concreteness in our analysis, we quantify the
uncertainty in the cosmic baryon density inferred from the CMB with
the likelihood function given in Ref.~\cite{dasi} by DASI+DMR.

If analyses of the CMB anisotropy measurements change and provide a
value for $\ob$ that is higher than that inferred from standard BBN,
then this could have been an indication for a model with large and
disparate neutrino degeneracy parameters known as degenerate
BBN~\cite{dbbn}.  However, if the favored large mixing angle (LMA)
neutrino mixing parameters of the solution to the solar neutrino problem
are verified ({\it e.g.}, by the KamLAND experiment~\cite{kamland}),
then synchronized neutrino flavor transformation in the early universe
stringently limits neutrino degeneracies~\cite{synch} and degenerate
BBN is no longer a rescue.

In order to precisely predict the abundance of $^4$He from standard
BBN for varying baryon density and neutrino number, Lopez and
Turner~\cite{lopezturner} included finite-temperature radiative,
Coulomb and finite-nucleon-mass corrections to the weak rates;
order-$\alpha$ quantum-electrodynamic correction to the plasma
density, electron mass, and neutrino temperature; and incomplete
neutrino decoupling.  Ref.~\cite{lopezturner} provides a fitting
formula for their results of the predicted helium abundance as a
function of $\eta$, $N_\nu$ and neutron lifetime $\tau$.  We employ
this fitting formula for the predicted $^4$He abundance, taking into
account the typographical correction of signs noted in
Ref.~\cite{Burles:2000zk}.  After all of the above corrections are
applied, the uncertainty in the predicted helium abundance is
dominated by the neutron lifetime uncertainty, which is now known
better than 0.1\%~\cite{rpp}.  Therefore, we can safely ignore the
theoretical errors, as they are dwarfed by observational uncertainty,
which we now address.

The primordial helium abundance $Y_p$ has been estimated in
observations of hydrogen and helium emission lines from regions of
hot, ionized metal-poor gas in dwarf galaxies (H{\sc ii} regions).  By
extrapolating the helium abundance and metallicity relationship for
these regions to zero metallicity, Olive, Steigman and
Skillman~\cite{oss}, and Fields and Olive~\cite{Fields:1998gv} find
\begin{equation}
Y_p(\text{OSS-FO}) = 0.238 \pm 0.002\ \text{(stat.)}\pm 0.005\
\text{(sys.)}\,,
\end{equation}
while Izotov and Thuan~\cite{it} find
\begin{equation}
Y_p({\rm IT}) = 0.244 \pm 0.002\ \text{(stat.)}\,.
\end{equation}
Uncertainties regarding the ionization structure and temperature
uniformity of the H{\sc ii} regions as well as underlying stellar
absorption are sources of significant systematic error.
Refs.~\cite{os,oss} estimate systematic effects in the primordial
helium abundance are 2\%.  Ref.~\cite{sauerjedamzik} finds that 
systematic effects can lead to 2-4\% uncertainties that tend to {\it
overestimate} $Y_p$.  In an attempt to avoid bias, in this work we
adopt the central value of
\begin{equation}
Y_p = 0.241 \pm 0.002\ \text{(stat.)}\pm \sigma_{\rm sys}\,,
\label{yp}
\end{equation} 
and characterize the systematic uncertainty as the disparity between
competing claims
\begin{equation}
\sigma_{\rm sys} = {|Y_p(\text{OSS-FO}) -Y_p({\rm IT})|}\,,
\end{equation}
or approximately 3\%.  The shape of systematic uncertainty in
likelihood space is certainly not well defined, therefore we make the
simplifying ansatz of a Gaussian distribution, as done, {\it e.g.}, in
Ref.~\cite{Cyburt:2001pq}, and combine the statistical and systematic
errors in quadrature.

\begin{figure}
\begin{center}
\includegraphics[width=9cm]{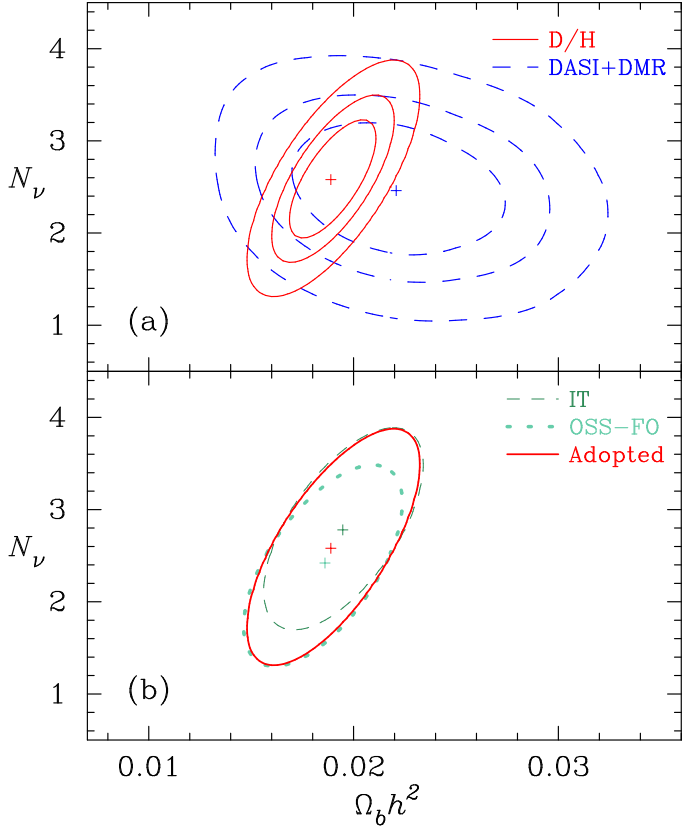}
\caption[]{\label{nnufull} 
Shown are the contours of 68\%, 95\% and
99\% CL (inner to outer contours) when using the baryon density
inferred from the deuterium abundance (D/H) at high-redshift (solid
lines) {\it or} that from the DASI+DMR analysis of CMB anisotropies
(dashed lines), in frame (a).  In both cases, we use our our adopted
determination of the primordial $^4$He abundance.  In frame (b), we
show the 99\% CL contours using the Izotov \& Thuan (IT), Olive,
Steigman \& Skillman (OSS) and Fields \& Olive (FO) and our adopted
value and uncertainty of the primordial $^4$He abundance along with
the D/H determination of $\ob$.  The uncertainty in $N_\nu$ does not
depend on the choice between IT and OSS-FO as much as the size of
systematic uncertainty.  See text for details.}
\end{center}
\end{figure}

We produce probability distribution functions (p.d.f.'s) for $N_\nu$
versus $\ob$, using Gaussian distributions for $Y_p$
[Eq.~(\ref{yp})] and $\ob({\rm D/H},N_\nu)$
[Eq.~(\ref{omegabdhnnu})], and the likelihood function given in
Ref.~\cite{dasi} for $\ob({\rm DASI})$.  We find, using
either the information from deuterium or the CMB on $\ob$:
\begin{eqnarray}
N_\nu{\rm(D/H)} &=& 2.60^{+0.90}_{-0.90} \\
N_\nu{\rm(DASI)} &=& 2.46^{+1.03}_{-1.01} \,,
\end{eqnarray}
at 95\% CL, which is consistent with the standard BBN prediction.  We
show the shapes of the likelihood contours in Fig.~\ref{nnufull}~(a).
To illustrate the difference between adopting the IT or OSS-FO helium
values, we plot the 99\% likelihood contours for the choices
$Y_p(\text{OSS-FO})$, $Y_p({\rm IT})$ (using the systematic
uncertainty of $\pm 0.005$) and our choice (\ref{yp}).  As seen in
Fig.~\ref{nnufull}~(b) the range of uncertainty does not depend on the
choice of the central value but the size of systematic effects.

\begin{figure}
\begin{center}
\includegraphics*[width=9cm]{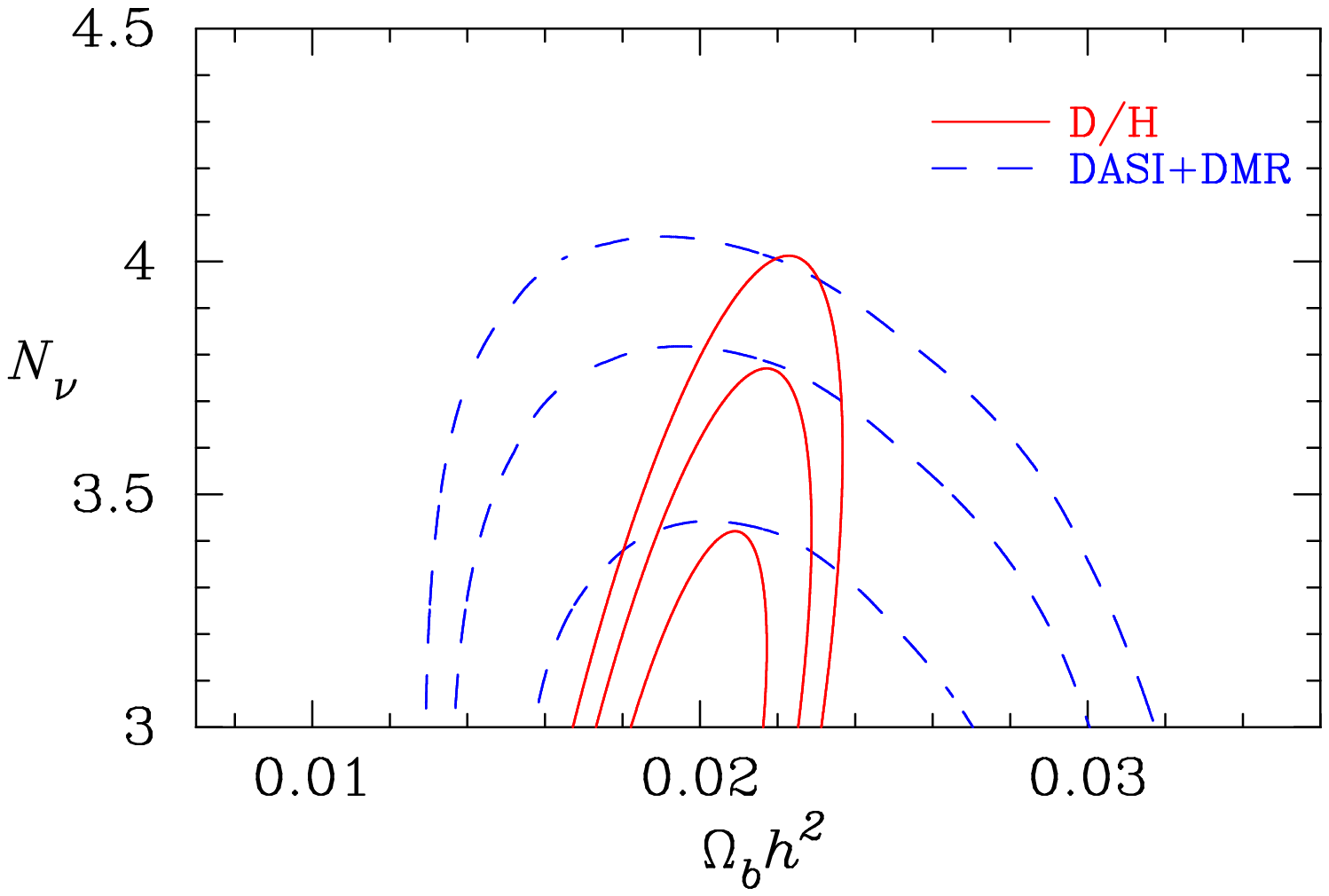}
\caption[]{\label{nnu}Shown are the contours of 68\%, 95\% and 99\% CL
  (inner to outer contours) given the prior condition that the active
  neutrinos are thermalized, our adopted value of primordial $^4$He
  abundance, and $\ob$ determined by D/H (solid lines) or the CMB
  observations by DASI+DMR (dashed lines).  See text for details.}
\end{center}
\end{figure}

For a sterile neutrino to be thermalized with the bath of the early
universe, the active neutrinos must be thermalized initially.  This
constitutes prior information that may {\it loosen} the constraints
shown in Fig.~\ref{nnufull}.  Prior information can be included in a
Bayesian approach~\cite{rpp}, integrating the p.d.f. only in the
physically allowed region, $N_\nu > 3$, which we have done using a
Monte Carlo integration.  We show the confidence level intervals for
this case in Fig.~\ref{nnu}.  As seen there, a fully populated fourth
neutrino is excluded at approximately the 99\% CL.

Measurements of CMB anisotropies by the MAP satellite
may measure $\ob$ to 10\%, giving a precise value independent of BBN.
If consistent with $\ob$ inferred from D/H and BBN, then $\ob$ becomes
a ``nuisance parameter'' that can be marginalized and the confidence
level for $N_\nu$ is simply the integral of the p.d.f.,
\begin{equation}
{\rm CL}(N_\nu) = \int_3^{N_\nu}{p(N_\nu^\prime|Y_p)\,dN_\nu^\prime}.
\end{equation}

\section{Four Neutrino Models in the Early Universe}
\label{neutrino}
There is now convincing evidence for neutrino flavor states to be
composed of large amplitudes of more than one mass state from two
experiments: Super-Kamiokande~\cite{superk} and the Sudbury Neutrino
Observatory (SNO)~\cite{sno}.  There is also an indication of a
neutrino oscillation signal at short baselines from the Liquid
Scintillator Neutrino Detector (LSND) experiment~\cite{lsnd}.  To
accommodate all three of these results, a four neutrino model must be
invoked (or $CPT$ is violated; see below).  The mass and
flavor state bases are related by a unitary
transformation
\begin{equation}
\nu_\alpha = \sum_i^4 U_{\alpha i} \nu_i \,,
\end{equation}
where $\alpha=e,\mu,\tau,s$ denotes the flavor state and $i$ is the
mass state.  The matrix $U_{\alpha i}$ generally has 6 rotation
(mixing) angles and 3 $CP$ violating phases.  The transformation
probability has the form:
\begin{align}
P(\nu_\alpha\rightarrow\nu_\beta) =\ &\delta_{\alpha\beta} - 4
  \sum_{i>j}{\rm Re}\left(U_{\alpha i}^\ast U_{\beta i} U_{\alpha j}
  U_{\beta j}^\ast\right)\ \sin^2 \left(\delta
  m^2_{ij}\frac{L}{4E}\right)\cr &+ 2 \sum_{i>j}{\rm
  Im}\left(U_{\alpha i}^\ast U_{\beta i} U_{\alpha j} U_{\beta
  j}^\ast\right)\ \sin \left(\delta m^2_{ij}\frac{L}{2E}\right),
\end{align}  
which provides rich neutrino oscillation phenomenology and
experimental possibilities~\cite{telling}.  Here $\delta m^2_{ij}
\equiv m_i^2 - m_j^2$, $L$ is the distance from where the flavor
$\nu_\alpha$ was created, and $E$ is its energy.  For reviews of
neutrino phenomenology see, {\it e.g.}, Refs.~\cite{nureview}.

For neutrino physics in the early universe, we are interested in the
magnitude of mixing amplitudes of active neutrinos converting to
sterile states.  First derived by Langacker~\cite{Langacker:1989sv},
and Dolgov and Barbieri~\cite{Barbieri:1989ti}, it has been known for
some time that a sterile neutrino coupling with a single active
neutrino via the unitary neutrino mass matrix must not have
~\cite{enqvist,shischrammfields}
\begin{equation}
\delta m_{\alpha s}^2 \sin^4 2\theta_{\rm BBN} \lesssim
\begin{cases}
5\times 10^{-6},\qquad\text{for $\alpha=e$} 
\\
3\times 10^{-6},\qquad\text{for $\alpha=\mu,\tau$}
\end{cases}
\,,
\label{oldconstraints}
\end{equation}
in order to not fully thermalize the sterile neutrino prior to BBN via
nonresonant collisional processes.  Such constraints
(\ref{oldconstraints}) certainly do not directly apply to multiple
neutrino mixing schemes including a sterile which nature may have
given us.  Multiple mixing angles and the phenomenon of lepton number
generation via neutrino mixing complicate the BBN bound.

We adopt a rotation angle ordering for $U$ so that
$(\nu_e,\nu_\mu,\nu_\tau,\nu_s) = (\nu_1,\nu_2,\nu_3,\nu_4)$ when all
mixing angles are set to zero.  Four neutrino mixing scheme
constraints have been examined recently in detail by Di
Bari~\cite{dibari}, which we summarize and expand on here in view of
the recent global analyses of four-neutrino models by Maltoni, Schwetz
and Valle~\cite{maltoni}.  The effective oscillation amplitude of
active-sterile neutrino mixing between flavor $\alpha$ and the sterile
can be written as
\begin{equation}
A_{\alpha;s} = 4|U_{\alpha 4}|^2 |U_{s 4}|^2 \simeq \sin^2
2\theta_{\rm BBN}\,.
\end{equation}
Therefore, BBN constraints for active-sterile mixing through a pair of
neutrino mass eigenstates $\nu_4,\nu_i$ are
\begin{equation}
\delta m_{4i}^2 A^2_{\alpha;s} \lesssim 
\begin{cases}
5\times 10^{-6},\qquad\text{for $\alpha=e$} 
\\
3\times 10^{-6},\qquad\text{for $\alpha=\mu,\tau$}
\end{cases}
\,,
\label{constraint}
\end{equation}
for nonresonant sterile production, where $\delta m_{4i}^2 >0$.  This
constraint (\ref{constraint}) applies primarily to the active-sterile
mixing that leads the oscillation, {\it i.e.}, that which has the
shortest oscillation length or largest $\delta m_{ij}^2$. Small vacuum
mixing amplitudes in the leading oscillation mode may avoid
thermalization, and so secondary oscillation modes with larger mixing
amplitudes may thermalize the sterile.

\subsection{3+1}
Models referred to as (3+1) may satisfy all experimental indications
of neutrino oscillations with a triplet of mass eigenstates that
provide the atmospheric and solar mass-scales, and a sterile-dominated
mass eigenstate with a large mass-scale splitting with the triplet
providing the LSND result via indirect $\bar\nu_\mu \rightarrow
\bar\nu_e$ mixing through the sterile.  In order to satisfy the mixing
amplitude that would provide the LSND signal, the oscillation
amplitude must be~\cite{maltoni}
\begin{eqnarray}
A_{\mu;e}\  &=&\  4 |U_{e4}|^2 |U_{\mu 4}|^2\cr
&>&\ 3 \times 10^{-4}\qquad ({\rm 99\%\ CL})\,,
\label{lsnd31}
\end{eqnarray}
at a $\delta m^2_{\rm LSND}\simeq 2\ {\rm eV}^2$ from Fig.~8 of
Ref.~\cite{maltoni}.  Indirect mixing of this form has two mixing
amplitudes that may thermalize the sterile.  In this case, the mixing
amplitudes $A_{\mu;s}$ and $A_{e;s}$ may participate in the
thermalization.  Consider the slightly less-constrained
[cf. (\ref{constraint})] $A_{\mu;s}= 4|U_{\mu 4}|^2 |U_{s 4}|^2$.
The mixing matrix element
\begin{equation}
|U_{s 4}|^2 > 0.54\quad ({\rm 99\%\ CL})\,,
\label{atm31}
\end{equation}
is bounded from below from constraints on the fraction of sterile
neutrinos participating in atmospheric oscillations (see Fig.~3 of
Ref.~\cite{maltoni}).

Evading constraints from BBN on $A_{\mu;s}$ and $A_{e;s}$ would
require minimizing both $|U_{e4}|$ and $|U_{\mu 4}|$ while satisfying
(\ref{lsnd31}).  This gives $|U_{e4}|^2 = |U_{\mu 4}|^2 \simeq
10^{-2}$.  Combined with the limit (\ref{atm31}), the BBN constrained
combination has the minimum value
\begin{equation}
\delta m^2_{\rm LSND} A^2_{\mu;s} \gtrsim 7\times 10^{-4}
\end{equation}
which exceeds the limits (\ref{constraint}) by at least two orders of
magnitude and invariably thermalizes the sterile.  This constraint
comes from the conservative case where $\delta m_{4i}^2 > 0$.  The
inverted case $\delta m_{4i}^2 < 0$ is resonant and more stringently
constrained.  Therefore, (3+1) models are strongly disfavored by
standard BBN.

\subsection{2+2}
\label{twoplustwo}
Four neutrino models may also accommodate all indications for neutrino
mixing with mass eigenstates in a pair of doublets that provide the
solar and atmospheric mass scales and a large mass gap between the
doublets providing the LSND mass scale.  The global analysis by
Maltoni {\it et al.}~\cite{maltoni} finds that such (2+2) models are
consistent within 99\% CL either with the sterile neutrino completely
participating in the atmospheric or solar solutions.  Therefore, it is
possible to choose a very small amplitude mixing between the doublet
not participating in sterile oscillations and the sterile state such
that the large LSND mass splitting does not populate the sterile
neutrino.  On the other hand, unitarity constrains the sterile state
to be present among some linear combination of mass eigenstates.
Whether (2+2) scenarios are compatible with the exclusion of large
sterile components in both the atmospheric and solar neutrino
observations is controversial~\cite{strumiabargerbahcall}.

Whether the sterile flavor is in the atmospheric doublet or solar
doublet is not of concern for the early universe.  In either case, the
sterile flavor will be thermalized.  For our notation, we employ the
mass scheme used in Ref.~\cite{maltoni}, where the (2+2) model has the
solar scale between $\nu_1$ and $\nu_4$ and the atmospheric scale
between $\nu_2$ and $\nu_3$, with the hierarchy only being determined
by the condition for resonance in the sun, $m_1<m_4$.  One can avoid
the effects of thermalization of the largest mass scale by setting the
inter-doublet mixings to zero, $U_{s1}=U_{s3}=0$, but by unitarity
$|U_{s2}|^2 + |U_{s4}|^2 = 1$, whereby the sterile neutrino
participates in large part in the solar scale, atmospheric scale or
both.  Having a complete sterile solution for either scale has already
been known to thermalize the sterile
neutrino~\cite{enqvist,shischrammfields}.  One could consider
democratically separating the sterile into both the atmospheric and
solar scales to minimize its presence in both, so that $|U_{s2}|^2 =
|U_{s4}|^2 =1/2$.  However, the oscillation amplitudes $A_{\mu;s} = 4
|U_{\mu 2}|^2 |U_{s 2}|^2$ and $A_{e;s} = 4 |U_{e 4}|^2 |U_{s 4}|^2$
still grossly exceed the BBN bounds~(\ref{constraint}) since the
magnitudes $|U_{e4}|^2 = |U_{e1}|^2 \tan^2\theta_{\rm LMA}$ and
$|U_{\mu 2}|$ must be large to accommodate the large to maximal mixing
angle solutions of the solar and atmospheric neutrino problems.
Therefore, (2+2) models are also strongly disfavored by standard BBN.

\subsection{Self-Suppression}

The possibility that a four neutrino mass scheme could be arranged in
such a way as to evade sterile-thermalization constraints were
considered by Bell, Foot \& Volkas~\cite{bell} and Shi, Fuller \&
Abazajian~\cite{shi}.  One could potentially either self-generate a
lepton number and suppress the large-mixing-amplitude thermalization
or offset the effects of sterile thermalization by altering the
electron neutrino-antineutrino asymmetry through alteration of
beta-equilibrium,
\begin{eqnarray}
{\rm n} + \nu_e\, &\leftrightarrow& {\rm p} + e^-\cr
{\rm n} + e^+&\leftrightarrow&{\rm p} + \bar\nu_e  \,.
\label{beta}
\end{eqnarray}
In the models considered in Refs.~\cite{bell,shi}, the direct
thermalization bounds~(\ref{constraint}) were avoided by placing the
sterile neutrino in the small-mixing-angle solution to the solar
neutrino problem, a region of parameter space still viable at the time
and outside of the constraint region~(\ref{constraint}).  

In addition, Refs.~\cite{bell,shi} explored methods of generating
asymmetries between electron neutrinos and antineutrinos by resonant
lepton number generation~\cite{leptasym}.  The resonance condition in
the early universe requires $m_4 < m_i$, where $m_i$ is a mass
eigenstate (more) closely associated with an active flavor.  A
positive electron neutrino number will suppress the $^4$He abundance
by shifting the rates~(\ref{beta}), which would be necessary if
standard BBN is inconsistent by having too high of a predicted $^4$He
abundance for a given $\ob$.  As shown above, standard BBN remains
consistent within observational uncertainty.  The sign of resonantly
generated electron neutrino/antineutrino asymmetry can be
chaotic~\cite{shichaos}, or at least not well
determined~\cite{chaosfinns}, having an significant chance (50\% if
chaotically random) of being negative and actually {\it increasing}
$Y_p$ by altering beta-equilibrium in the opposite direction.  If the
sign of the asymmetry {\it is} randomly chaotic, then causally
disconnected regions will have different sign asymmetries, which leads
to an enhancement of the transformation of active neutrinos into
sterile neutrinos at the boundaries of regions of different
sign~\cite{shifuller} and potentially placing more stringent
constraints on four neutrino mass schemes~\cite{abazajianfullershi}.

There is considerable evidence now that the solar solution lies in the
LMA region of parameter space~\cite{sno}.
Therefore, as discussed in the previous sections, thermalization of
the sterile is unavoidable in either the (3+1) or (2+2) scenarios.
And, importantly, it was shown by Di Bari~\cite{dibari} that
thermalization of the sterile in these four neutrino models suppresses
lepton number generation, and electron neutrino/antineutrino
asymmetries are not effective in avoiding the BBN
bounds~(\ref{constraint}).

\section{Constraint Evasion and New Physics}
\label{caveat}
The simplifying and appealing principle of Occam's razor has proven to
be a powerful tool as a predictor in science, yet nature does not
always take the most simple form.  The minimal model for four neutrino
mixing or the standard BBN described above may certainly not be the
entire framework of the early universe or particle physics.
Importantly, if all experimental indications for neutrino oscillations
remain, {\it viz.}, if the MiniBooNE detector~\cite{boone} verifies
the LSND signal, K2K~\cite{k2k} and MINOS~\cite{minos} verify the
atmospheric oscillation solution and KamLAND detects the LMA
signal~\cite{kamland}, then new physics must be at play beyond
standard three-neutrino mixing and standard BBN.  There exist a number
of ways of accommodating such a scenario, several of which are
described below.  The aesthetic value of these scenarios are left to
the judgment of the reader.

{\it Pre-existing lepton asymmetry}  ---  A lepton number in the
active neutrino flavors will suppress sterile neutrino population by
magnifying the associated lepton potential and dwarfing the vacuum
mixing amplitude~\cite{leptasymsupp}.  This lepton number would have
to be produced by an unspecified mechanism earlier than the population
of the sterile neutrino would take place.

{\it A fifth mass eigenstate} --- Appropriate insertion of a mass
eigenstate with a major sterile component with $m_5 < m_i\ (i=1...4)$ in
degenerate neutrino mass models may resonantly generate lepton number
sufficiently prior to sterile thermalization as to suppress it.  This
possibility was explored in Ref.~\cite{dibari}.

{\it Majoron fields} --- One mechanism for generating neutrino mass
involves a massless Nambu-Goldstone boson (a majoron) from models
where either the total or partial lepton number is spontaneously
broken~\cite{Chikashige:1980ui}.  In such models, a coherent majoron
field creates potentials for the neutrinos proportional to the
gradient of the field and suppresses sterile thermalization in a
similar way as a pre-existing lepton asymmetry~\cite{Bento:2001xi}.
Interestingly, this mechanism arises from the neutrino mass model
itself.

{\it A low reheating temperature universe} --- There is no direct
evidence that the neutrino background is thermalized.  As explored in
Refs.~\cite{lowrh}, the highest temperature of the universe could have
been only $0.7\rm\ MeV$.  The neutrino background may never have been
thermalized, but the observed light element abundances could still be
created.  In this case, sterile neutrinos may modify the
nucleosynthesis processes by partial population but are not directly
excluded.

{\it Baryon-antibaryon inhomogeneity} --- Detailed calculations of
diffusion and nucleosynthesis in universes containing baryon number
asymmetries~\cite{Steigman:ev} have found that small-scale antibaryon
domains are not excluded by BBN and the observed light element
abundances~\cite{antibbn}, and may lift constraints on relativistic
energy density present at BBN to $N_\nu \lesssim 7$, even with total
baryon densities consistent with CMB
observations~\cite{Giovannini:2002qw}.

{\it Extended quintessence} --- Non-minimally coupled quintessence
models (where the quintessence field is not only coupled to gravity)
that provide a negative-pressure vacuum energy density to explain the
acceleration phase that the universe may be entering can alter
BBN~\cite{Chen:2000xx}.  In certain cases of such ``extended
quintessence'' scenarios, the quintessence field may behave to
decrease the expansion rate during the freeze-out of beta
equilibrium~(\ref{beta}), and therefore {\it decrease} the predicted
helium abundance.  This reduction of the expansion rate could offset
the increase in the expansion rate due to the presence of an extra
neutrino degree of freedom and allow for four-neutrino models.

{\it $CPT$ violating neutrinos} --- There exists a radical proposal
that fits all indications for neutrino oscillation and invokes $CPT$
violation in the neutrino sector~\cite{nocpt}.  The success of this
model lies in the fact that LSND's indication for neutrino oscillation
lies primarily in the antineutrino $\bar\nu_\mu \rightarrow\bar\nu_e$
channel~\cite{lsnd}, is motivated by braneworld scenarios with extra
dimensions, and gives dramatic predictions for the
MiniBooNE~\cite{boone} and KamLAND~\cite{kamland} experiments.  This
model has no effect on standard BBN since in the standard case the
neutrinos and antineutrinos are equally thermally populated and
therefore $CPT$ violating neutrino oscillations do not disturb the
detailed balance of thermal equilibrium, leading to no direct
conflicts between light element abundances and standard BBN.

\section{Discussion and Conclusions}

In minimal models of big bang nucleosynthesis with no new physics, we
have shown that four neutrino models explaining current indications
for neutrino oscillations are disfavored at the 99\% CL.  This
conclusion depends on systematic effects not being larger than that
expected ($\sim 3\%$) in determining the $^4$He abundance in ionized
H{\sc ii} regions and that the baryon density inferred by the D/H
abundance in six high-redshift quasar absorption systems and the
anisotropies in the cosmic microwave background are approaching the
true cosmic value of $\ob$.  

The MAP satellite will verify or disprove the value of $\ob$ inferred
by the above methods to high precision in the near future~\cite{map}.  If
consistency remains in cosmological determinations of the baryon
density, then $\ob$ would then become a ``nuisance'' parameter in
determining the cosmological energy density at standard BBN.  The
primordial helium abundance is currently the best probe of the energy
density of the universe present at BBN, yet there remain no concrete
proposals in the literature for the reducing systematic uncertainties
present in determining the primordial $^4$He abundance, which is the
dominant uncertainty in constraining the energy density present in the
universe at the age of one second via standard BBN.

In addition, we have summarized several scenarios that evade the
standard BBN model constraints presented here.  Remarkably, if all
experimental indications for neutrino oscillations are confirmed, new
physics must be present not only in the particle content of the
neutrino sector but also in the early universe.

\section{Acknowledgments}

I would like to thank Kaladi Babu, Gabriela Barenboim, Nicole Bell,
Scott Burles, Janet Conrad, Scott Dodelson, Josh Frieman, George
Fuller, Manoj Kaplinghat, Jim Kneller, Rabi Mohapatra, Sandip Pakvasa,
Subir Sarkar, Mike Turner and Jose Valle for fruitful discussions, and
the Institute for Nuclear Theory at the University of Washington for
hospitality and the DOE for support in hosting a Mini-Workshop on
Neutrino Masses \& Mixing which initiated this project.  I would
especially like to thank John Beacom for extremely valuable
discussions regarding my statistical approach.  This research was
supported by the DOE and NASA grant NAG 5-10842 at Fermilab.

\end{document}